\documentclass[prl,twocolumn,preprintnumbers,amsmath,amssymb,showpacs]{revtex4}
\usepackage{graphicx}
\begin{document}

\title[]{Secure Quantum Telephones}

\author{J.S. Shaari$^{\dag}$}
\author{R.S. Said$^{\dag\ddag}$}
\author{M.R.B. Wahiddin$^{\dag\ddag}$}
\affiliation{$^{\dag}$Faculty of Science, International Islamic University Malaysia, Kuantan, 25710 Pahang, Malaysia. \\$^{\ddag}$Information Security Group, MIMOS Berhad. Technology Park Malaysia, 57000 Kuala Lumpur, Malaysia.}

\begin{abstract}
We propose in this paper a novel deterministic protocol using particular maximally entangled states of polarized photons for a genuine bidirectional secure communication setup. We further propose a plausible experimental setup for such a protocol using currently available optical technology which liberates two communicating party from sharing identical apparatus. We note that security of the protocol is promised by the monogamous nature of maximally entangled states. We discuss the security of our protocol in the light of some simple eavesdropping scenarios.
\end{abstract}

\maketitle

{\it Introduction.}
Quantum key distribution (QKD) is arguably the only currently feasible realization of the application of quantum mechanics to the field of information theory and cryptography. Promising security based on the postulates of quantum mechanics, it allows one to abandon the notion of security in computational complexity, e.g. the RSA \cite{RSA78} and by making no presuppositions regarding an adversary Eve save that she functions only in the possibilities of the quantum world. A host of QKD protocols emerged starting from the pioneering work of BB84 \cite{bb84}, entanglement based protocols like Ekert 91 \cite{ekert} and variants of the two including \cite{GBB84,durt} to name only a few. Deterministic setups then came to be proposed in principle for the possibility of secure direct communication (we note that thus far, this would only be true in the context of a noiseless channel). Amongst them were reported in \cite{Bos02,Cai04,Luc05,qdial,6dp}. Other protocols like Chen {\it et al.} \cite{ent3d} includes deterministic only in the context of no wastage of qubits due to wrong measurement bases or even in the test for an eavesdropper.
However, most of these protocols mentioned (excepting \cite{qdial}) do not allow for a bidirectional form of communication without necessitating double setups; or simply put, for any such setup allowing Alice to encode bits for Bob, another duplicate to allow for Bob to encode bits for Alice is necessary.
We propose in this paper a novel protocol making use of two nonsinglet maximally entangled states to allow for such a bidirectional communication to be achieved securely; namely `secure quantum telephone'. We adopt the definition of deterministic (as proposed in \cite{6dp}), when the encoding/decoding procedure in principle allows a recipient to infer with certainty what was encoded by the transmitter.
We note a protocol for quantum dialogue was reported in \cite{qdial}. However, the protocol carries a flaw where a Quantum Man in the Middle (QMM) attack would go totally undetected. Another work \cite{GHZ} rescues the flaw and was presented using the GHZ states. However in contrast, our proposal is certainly more feasible as we do not resort to Bell measurements nor GHZ state measurements. We further propose a possible experimental setup which may be used to realise the protocol with currently available technology. We briefly also discuss a possible modification to our protocol to improve its efficiency. The security of our protocol is demonstrated by considering an Eve attacking using a simple Intercept Resend (IR) strategy  with von Neumann measurements as well as the QMM attacks.

{\it The protocol.}
We make use of two Bell states that can be generated through the process of spontaneous parametric down conversion in nonlinear crystal,
\begin{eqnarray}
|\psi^+\rangle &=& \frac{1}{\sqrt{2}}\left(|H\rangle|V\rangle+|V\rangle|H\rangle\right),\cr
|\phi^-\rangle &=& \frac{1}{\sqrt{2}}\left(|H\rangle|H\rangle-|V\rangle|V\rangle\right). \label{Bell}
\end{eqnarray}
The state $|H\rangle$ ($|V\rangle$) denotes horizontal (vertical) polarization state of photon. We describe the protocol as follows: Bob selects between the two possible Bell pairs, $|\psi^+\rangle$ and $|\phi^-\rangle$, and sends half of the pair to Alice who will measure in either the basis $\sigma_x=|H\rangle\langle V|+|V\rangle\langle H|$ or $\sigma_z=|H\rangle\langle H|-|V\rangle\langle V|$. After announcing her measurement (not results), Bob sends to Alice the other half and Alice again measures in the same basis as before. Hence, we can find the characteristic equations as:
\begin{eqnarray}
\sigma_x \otimes \sigma_x |\psi^+\rangle &=& |\psi^+\rangle, \cr
\sigma_z \otimes \sigma_z |\psi^+\rangle &=& -|\psi^+\rangle, \cr
\sigma_x \otimes \sigma_x |\phi^-\rangle &=& -|\phi^-\rangle, \cr
\sigma_z \otimes \sigma_z |\phi^-\rangle &=& |\phi^-\rangle.
\end{eqnarray}
 Alice then informs over an authenticated public channel whether the results were correlated or otherwise. Given equation (2), Bob would be able to know which basis Alice had chosen and they may then share a secret bit. Alice on the other hand, given her bases of choice, may know for certain the pair that Bob had sent. In this way, for every encoding done by Alice (choice of basis), Bob may decode it without ambiguity and likewise for every bit of Bob's encoding (choice of state), Alice would decode perfectly. We however note though that a more reasonable setup would have Alice and Bob encoding in separate runs (which is more sensible in a communication setup), i.e Alice or Bob's encoding run separately. To ensure security, Alice (Bob) would select (predetermined selection) qubit pairs to be tested for a CHSH \cite{CHSH} violation by making measurements in relevant bases to yield maximal violation. We note that this is reminiscent of the control modes in \cite{Bos02,Cai04,Luc05,6dp,qdial}. The test for a maximal CHSH violation would tell the communicating parties of the purity of their Bell states. It is well known that an entangled Bell pair exhibits a characteristic known as monogamy \cite{mono,mono1} where they do not allow another party to correlate with.

The control mode may be carried out as follows: Alice selects at random, with probability $C$ which of the pairs that she would use for a CHSH check and after her measurement on the first qubit, she informs Bob of the fact that it is a control mode. Bob then would proceed with his own measurements and such a process is sufficiently repeated with certain probability so that the CHSH value may be estimated. We note that Eve would only know of the control mode after Alice's measurement. We denote Alice's and Bob's bases measurement for the purpose of CHSH testing by $a_1^1$, $a_1^2$, $a_2^1$, and $a_2^2$. The CHSH parameter $S$, is given by
\begin{eqnarray}
S=E(a_1^1,a_2^1)-E(a_1^1,a_2^2)+E(a_1^2,a_2^1)+E(a_1^2,a_2^2)
\end{eqnarray}
and is equal to $2\sqrt{2}$ (or $-2\sqrt{2}$) for maximally entangled pure states. Hence, a maximal violation tells Alice that the channel is secure. We note that in fact in the instances of Bob's encoding run, a control mode may be requested by Bob after Alice's first measurement. The control modes are done where the encoding party has the freedom to choose which run would be for control mode and which for message. This is in contrast to \cite{qdial} where control mode is Alice's prerogative alone.

{\it Experimental Setup.}
As depicted in Figure 1, we propose a feasible free-space setup to implement our key distribution protocol. A laser pulse pumped into a barium borate (BBO) crystal will generate one of the four Bell states depending on the cutting angle of the crystal. Bob then prepares his Bell states, either $|\psi^+\rangle$ or $|\phi^-\rangle$  by controlling an electro-optical liquid crystal polarization (LCP) rotator, R1. After traversing through a fiber path {\it a-c}, one photon (half of the Bell pair) will be sent to Alice's lab while the other will be sent to the looped fiber L1 on the arm {\it b-c}. We require here polarization maintained (PM) fibers to implement the paths {\it a-c} and {\it b-c}. The length of loop L1 should be greater than the line-of-sight distance between Bob's lab and Alice's lab since the second photon can only be sent out from Bob's lab after a measurement has been done by Alice onto the first photon. Alice's lab is equipped with a LCP rotator R2, a polarized beam splitter PBS, and two avalanche photo diode detectors. Alice will choose the measurement basis by controlling the rotator R2. It is also important to consider the losses of the fiber since we need to delay the traveling time of photon 2 in fiber loop L1.
\begin{figure}[t]
	\centering
		\includegraphics[scale=0.815]{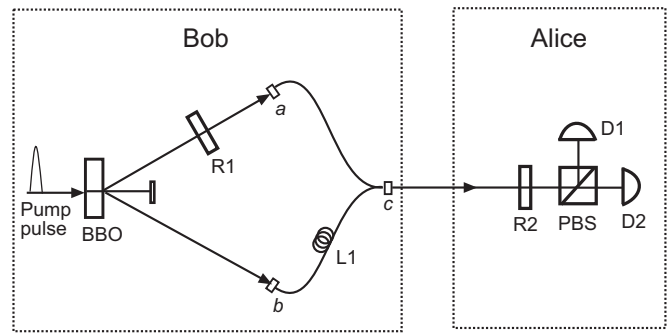}
	\label{fig6}
	\caption{A proposed implementation of the protocol.}
\end{figure}

{\it Eavesdropping and efficiency.}
In the simplest scenario where an adversary Eve attempts an Intercept Resend (IR) strategy, she needs to measure Bob's qubits before sending them to Alice thereafter. Such an attack would result in her sending a separable state to Alice and a CHSH test would give $|S|\leq 2$. Another method of attack conceivable by Eve would be the QMM where she would exchange the pairs she receives with her own Bell pair. Simply enough, she receives and distributes the relevant pairs in a sequential manner just as Bob would. In this way, Eve would know perfectly the information encoded by Alice and upon performing a Bell measurement to distinguish between the pure Bell states Bob sent, she would know what encoding Bob would deduce from Alice's announcement. However, in a control mode, Eve would have Alice measure her half of the Bell pair and her only way of avoiding detection is to somehow influence Bob's result so that Alice and Bob would have a maximal CHSH violation. To commit to a measurement herself on the qubit received from Bob would, uninterestingly translates into Alice and Bob again measuring noncorrelated separable states, resulting with $|S|\leq 2$. Having access to Alice's other half (disentangled state), say $|E\rangle$ as well as her half of Bob's entangled qubit her hope would be for Bob to make his measurements on the $|E\rangle$ (this implicitly suggests Eve's hope that she uses the same entangled state that Bob does). She therefore looks for a process $T$ as such
\begin{eqnarray}
T(|\Phi\rangle \langle \Phi|,|E\rangle)\rightarrow |E\rangle,|\chi\rangle \langle \chi|
\end{eqnarray}
The $|\Phi\rangle \langle \Phi|$ is the density operator of the entangled qubit at Bob's station and $|\chi\rangle \langle \chi|$ is the state that Eve has after the process $T$. The first term on the right hand side of equation(4) would be the state of the qubit now belonging to Bob and the second to Eve. We believe Eve may best achieve this by virtue of a teleportation scheme. However, differently from the conventional teleportation scheme, Bob would not be making any unitary transformation on his qubit and therefore only in half the instances would he actually make a measurement on $|E\rangle$. The other half the time, Bob would be making a measurement on a state orthogonal to $|E\rangle$. We may imagine then Alice and Bob testing CHSH on the a state that is a mixture of the entangled states expected.
\begin{eqnarray}
\rho=\frac{1}{2}|\psi^+\rangle \langle \psi^+| +\frac{1}{2}|\phi^-\rangle \langle \phi^-|
\end{eqnarray}
and $S(\rho)=0$.

Alternatively, Alice and Bob may check for errors, $d$ instead of the CHSH and for the IR and QMM attack, the probability of detection gives $d=25\%$ and $d=50\%$ respectively. Eve's probability of stealing say, $I=nI_{AE}$ bit of information without being detected is $P(C,d)=(1-C)^n/[1-C(1-d)]^n$ \cite{Bos02} where $I_{AE}$ is the amount of mutual information shared between Alice and Eve given an attack scheme. It is straightforward to see for $c>0$, the protocol is asymptotically secure.

The theoretical efficiency of a QKD protocol as given by Cabello \cite{Cab00} is expressed as $\mathcal{E}$ $=
b_{s}/(q_{t}+b_{t})$, where $b_{s}$ is the expected number of secret bits received, $q_{t}$ the number of transmitted qubits on the quantum channel, and $b_{t}$ the number of transmitted bits on the classical channel. It is straightforward to calculate our protocol as having an efficiency of $\mathcal{E}=[2/(2+2)]=1/2$. However, as mentioned above, it is more reasonable to assume that Alice and Bob would encode/ decode in separate runs. Hence, an average efficiency for the protocol is $\mathcal{E}_min=[1/(2+1)+1/(2+2)]/2=7/24$ (Alice's encoding differs from Bob's). This would admittedly be comparable to that of BB84. We further note that we do not consider control modes in our calculations in the spirit of \cite{Cab00}, where the probability $c$ may be taken to be small and therefore the bits sacrificed are negligible.

{\it Modification and increased efficiency.}
We now deliberate on a possible modification of our protocol to increase the efficiency and essentially improves the work in \cite{qdial}. In a certain sense, it is essentially very much similar to the protocol reported in \cite{QSDC} with an additional step allowing for either of the parties to encode/decode information. We consider Bob sending any of the four Bell states in a sequential manner as before and the encodings would be in the Bell states, i.e, $\left\vert\psi^+\right\rangle\equiv 00$,$\left\vert\psi^-\right\rangle\equiv 01$,$\left\vert\phi^+\right\rangle\equiv 10$ and $\left\vert\phi^-\right\rangle\equiv 11$. We note that in \cite{QSDC}, encoding is reflected in the unitary operations producing these Bell states starting from the singlet state $\left\vert\psi^-\right\rangle$. Alice upon receiving Bob's qubit would acknowledge receipt over the public channel. Bob then sends the second qubit and Alice makes Bell measurements to distinguish which of the Bell states Bob sent. Alice then considers a unitary transformation (either of the three Pauli matrices, $\sigma_{1},\sigma_{2},\sigma_{3}$ or the identity operation, $\sigma_{0}=I$) which would transform the state that Bob sent;
\begin{eqnarray}
\sigma_{i}\otimes I\left\vert \Psi\right\rangle  & \rightarrow\left\vert \Psi^{\prime
}\right\rangle
i=0,1,2,3, \left\vert \Psi\right\rangle,\left\vert \Psi^{\prime}\right\rangle \in\left\{\psi^{\pm},\phi^{\pm}\right\}.
\end{eqnarray}
In order to encode and send the desired bits to Bob as $\left\vert \Psi^{\prime}\right\rangle$,
Alice may just inform Bob over the public channel of the relevant operation considered. Bob would deduce with certainty Alice's encoding as the resulting state of the operation on the state he sent. A control mode would proceed with Alice making projective measurements of the first qubit and informs Bob of the basis used and result to check for errors. We further note that the efficiency of the modified protocol $\mathcal{E}_m$ is essentially $\mathcal{E}_m=4/(2+3)=4/5$ (average=$[2/(2+3)+2/(2+1)]/2=8/15$) where the three classical bits come from Alice's receipt and operation disclosed.
However, all in all, given the experimental demands of this modification which includes Bell measurements as well as more quantum memory capacity (fiber loop on both Alice and Bob's side), we prefer to think our first protocol as a more proper practical solution for a secure quantum telephone.

{\it Conclusion.}
We have presented in this paper a novel and simple protocol for two way secure quantum communication. In line with simplicity of a protocol and ease of possible implementation, our work does not require Bell measurements, rather only sharp von Neumann measurements and security is promised by the monogamous nature of maximally entangled pairs. This is checked by testing the CHSH violation to ensure purity of the states sent. The theoretical efficiency of our protocol is comparable to BB84. While granting security in the face of an IR attack, we note that our protocol does not share the flaw against QMM attack in \cite{qdial}.
Our proposed optical setup is very much feasible given current technology.
We note that in reality, the protocol works as a `telephone' so to speak only in the case of noiseless channels and in the case otherwise, the protocol may be used (supplemented by error correction schemes and privacy amplification) for a normal QKD setup.

We would like to thank Suryadi and Suhairi Saharudin at MIMOS Berhad for fruitful discussions.

\end{document}